# Lasing and spasing with active individual core-shell plasmonic nanoresonators

*Dávid Vass[1,2], András Szenes[1,2], Péter Zoltán Nagy[1], Balázs Bánhelyi[2,3] and Mária Csete[1,2*]*

[1]*University of Szeged, Department of Optics and Quantum Electronics, Szeged 6720, Hungary*
[2]*Wigner Research Centre for Physics, Budapest 1121, Hungary*
[3]*University of Szeged, Department of Computational Optimization, Szeged 6720, Hungary*
*\*mcsete@physx.u-szeged.hu*

**Abstract**

Active core-shell nanoresonators were designed in order to achieve large near-field enhancement, large power-outflow and minimal spaser threshold in the pump **E**-field strength. Gain-metal-dielectric (GMD) and gain-metal-gain (GMG) nanoresonator compositions were optimized with corresponding objective functions. The average local **E**-field, power-outflow and extinction cross-section were mapped above the pump **E**-field strength and dye concentration parameter plane with the criterion that the local **E**-field is smaller than the damage threshold of the nanoresonator. Regions, corresponding to the maxima in the average local **E**-field, the highest power-outflow, or to the zero-crossing of the extinction cross-section, were selected for detailed studies. The spectral distribution of the near-field enhancement, optical cross-sections, optical responses, quantum efficiencies, as well as the polar angle distribution of the far-field radiated power and the local charge distribution of dominant modes were inspected. Based on the results the GMD nanoresonator composition is proposed to maximize local **E**-field in near-field amplifiers, to maximize power-outflow in far-field-emitting lasers and to minimize threshold **E**-field in spasers. Comparing the complete characteristics, both compositions are suitable for different operation regions, the GMG is proposed as near-field amplifier and far-field out-coupling nanolaser, whereas the GMD is unambiguously preferable to achieve optimal spaser properties.

**Keywords**: nanoresonator; lasing; spaser; numerical calculations; plasmon; optimization

## 1. Introduction

Core-shell (CS) nanoresonators (NRs) can support localized surface plasmon resonances (LSPRs), which are excited at the dielectric-metal interfaces due to the hybridization of elementary sphere and cavity plasmons [1]. The strength of the interaction between these plasmons is determined by the thickness of the metal shell. The plasmon resonance frequency can be tuned across a wide spectral range, from the visible to the infrared region, by varying the relative size of the inner ($r_1$) and outer ($r_2$) radii, thereby modifying the so-called generalized aspect ratio (*GAR=$r_1$/$r_2$*), as well as by adjusting the dielectric properties of the core, shell and the embedding medium [2, 3, 4]. Two distinct resonant GARs exist for CS nanoresonators at the same wavelength: a thin-shell geometry that promotes absorption and allows for larger near-field enhancement (*NFE*), and a thick-shell geometry that enhances scattering more efficiently [5]. Thin shell CSs have typically narrow-band absorption spectra, while the thick shell CSs exhibit broad-band scattering spectra.

Different types of nanoresonators can enhance the spontaneous emission of nearby emitters due to the large local density of states (LDOS), which stems from the large near-field enhancement [6, 7]. Plasmonic nanoparticles can also enhance the stimulated emission of nearby dye molecules, which is suitable to achieve and improve nanoscale lasing [8].

Numerical methods have already been used to model lasing media containing four-level laser dyes both in frequency and in time-domain. By separating the fields into constituents oscillating at the excitation and emission frequencies, the computational time can be significantly reduced [9]. Steady-state rather than time-domain calculations are reasonable to obtain the optical response of the system, when the external pulse duration is much longer than the time-scale required to reach stationary population densities. In this case, the steady-state population inversion can be derived from the rate equations [10]. In the frequency-domain, the dielectric properties of the gain medium can be described using Lorentzian permittivity both at the pump and the probe wavelengths, considering the inhomogeneity of the local **E**-field as well [11].

In plasmonic nanolasers, resistive losses occur in the metal, in addition to the non-radiative loss arising as a negative absorption in the gain material and the radiation losses related to lasing [12]. Effective **E**-field confinement is crucial in several applications to achieve an improved optical gain and reduced losses in nanolasers. Theoretical and experimental results demonstrate that semiconductor-metal core-shell nanoresonators can efficiently confine the **E**-field in three-dimensions, resulting in lower thresholds compared to their photonic nanolaser counterparts [13]. Nanolasers based on hybrid waveguide structures can also result in strong **E**-field confinement, in addition to the increased propagation distance inside the metallic waveguide, and the reduced gain threshold [14].

A special type of nanolaser is the spaser, which exhibits a fundamentally different operation characteristic, as instead of the photonic modes, plasmonic modes are at play. Experimental work has demonstrated the stimulated emission of surface plasmons from spasers [15]. Spasers can be constructed with different core-shell nanoresonator compositions. The spaser threshold can be expressed as the gain level at which the extinction cross-section (*ECS*) switches from a positive to a negative value [16]. Due to the dispersion of the material dielectric functions, the lasing threshold exhibits a dispersion [17]. Consequently, a minimal spaser threshold exists for small nanoparticles, which depends solely on the dielectric function of the metal and the gain material's dielectric properties. It was demonstrated that numerical modelling has the potential to predict threshold gain levels, but it can yield unphysical results in the post-threshold operation regions. To eliminate these numerical artifacts, a non-linear term has to be incorporated into the dielectric function of the gain [18].

To ensure comparability with lasing, the spasing threshold can be also expressed as a minimum in the pump **E**-field strength, but usually the threshold is specified in terms of the gain level required to achieve zero-crossing in *ECS* [19]. However, optically pumped spasers can also be out-coupled into the far-field above configuration specific thresholds, with an efficiency comparable to that of absorption in the metal.

In our previous studies we have shown that at a fixed dye concentration different operational regions can be reached by optimizing the geometry to achieve maximal near-field and maximal far-field out-coupling, on demand. By increasing the dye concentration in a near-field optimized nanolaser, the spasing operation region can be achieved [20].

Resonant gain (RG) phenomena can occur in multishell nanostructures by varying the concentration of dye molecules. As in the epsilon-near-zero-and-pole (ENP) region in 1D stacked multilayer, a huge amplification of the emitted photons can be achieved inside a multishell, when the self-enhanced loss compensation criterion is met, which makes the system preferable for high intensity spasing devices [21].

Plasmonically coupled spaser can be created using gold nanospheres on a gold film, surrounded by a dielectric layer doped with dye molecules. The gain threshold can be significantly reduced compared to a stand-alone gold nanosphere, as the coupled modes produce larger electric field, and the gain more efficiently compensates the metal loss [22].

A topological spaser can also be realized using a honeycomb lattice of spherical metal nanoshells, with a gain medium inside the cores. Topological spasers are capable of generating chiral plasmonic modes, and the spontaneous violation of the symmetry determines, which mode is at play. These phenomena can be used for ultrafast optical signal tailoring, biomedical sensing and detection [23].

Plasmonic lasing systems can be used in wide range of application areas. High intensity can be achieved e.g through lasing enhanced by random gold nanoparticles, or symmetric and asymmetric double grating plasmonic structures [24, 25]. Lasing-enhanced surface plasmon resonance (LESPR) sensors with improved sensing capabilities can be developed for refractive index sensing or gas detection [26]. Plasmonic lasing systems also hold promise in cancer therapy such as plasmonic nanostars surrounded by different dye molecules [27]. In our previous work, we have shown that dye molecules are advantageous to achieve uniform energy deposition inside extended targets, which is important in realizing plasmon-enhanced fusion [28].

In the present study two core-shell nanoresonator compositions were optimized to achieve large near-field enhancement, efficient power-outflow into the far-field, and minimal spaser threshold in the pump **E**-field strength corresponding to the zero-crossing of the extinction cross-section.

## 2. Methods

Numerical computations were performed in frequency domain using the finite element method (FEM) with the RF module of COMSOL Multiphysics, as described in our previous publication [20]. Initially, we optimized the geometry of dye doped dielectric-gold-dielectric core-shell nanoresonators of two different compositions: a gain-metal-dielectric (GMD) and a gain-metal-gain (GMG), using an in-house developed optimization algorithm (Fig. S1, Table S1) [29]. The gain medium consisted of a spherical dielectric core (with refractive index $n_{host}$ = 1.62 [10, 11]) seeded with four-level Rh800 dye molecules, characterized by absorbing and emitting oscillators at 680 nm and 710 nm, respectively (Table S1). These cores were covered by 5 nm thick gold shells in both nanoresonators, and the outer layer was a passive thin dielectric shell in GMD, while a slightly thicker active dye-doped dielectric shell, analogous with that of the gain medium in the core, covered the GMG NR.

Numerical pump-and-probe simulation was also conducted in the frequency domain. First, the steady-state population differences were determined using a monochromatic continuous-wave (CW) pump at the absorption wavelength of the dye. Subsequently, the pump was turned off and a low-intensity CW probe was injected at the emission wavelength, and the stimulated emission enhancement was determined self-consistently.

The steady-state population differences can be described as follows [11]:

$$\Delta N_a = N_3 - N_0 = \frac{\tau_{32}\Gamma_{03} - 1}{1 + (\tau_{32} + \tau_{21} + \tau_{10})\Gamma_{03}} N, \quad (1)$$

$$\Delta N_e = N_2 - N_1 = \frac{(\tau_{21} - \tau_{10})\Gamma_{03}}{1 + (\tau_{32} + \tau_{21} + \tau_{10})\Gamma_{03}} N, \quad (2)$$

where *N* is the total number density of the dye molecules, $N_0$, $N_1$, $N_2$ and $N_3$ are the number density of dye molecules at the different levels, $\tau_{32}$, $\tau_{21}$, $\tau_{10}$ are the lifetimes of the transitions ($\tau_{21} \gg \tau_{32}, \tau_{10}$). The $\Gamma_{03}$ is the steady-state, local **E**-field dependent effective probability of the 0-to-3 level transition, calculated as follows:

$$\Gamma_{03} = \frac{e^2 d_a^2 |\overrightarrow{E_{local}}|^2}{\hbar \Delta \omega_a}, \quad (3)$$

where *e* is the electron charge, $d_a$ is the transitional dipole moment, $E_{local}$ is the local **E**-field, $\hbar$ is the reduced Planck's constant and $\Delta\omega_a$ is the *FWHM* of the absorption line.

The local **E**-field dependent nonlinear term was included into the equations describing the Lorentzian spectral responses, in order to avoid numerical divergences at the probe wavelength [18, 19].

Based on the local **E**-field dependent population differences, the complex permittivity of the gain medium can be defined around the absorption and emission frequency, with the following equation:

$$\varepsilon(\omega) = \varepsilon_{host} + \frac{1}{\varepsilon_0} \sum_{j=a,e} \frac{\sigma_j \Delta N_j}{\omega^2 + i\omega\Delta\omega_j - \omega_j^2}, \qquad (4)$$

where j=a refers to the absorption and j=e to the emission, $\sigma_j$ is the coupling constant, $\Delta\omega_j$ is the *FWHM* of the corresponding spectral lines and $\omega_j$ is the frequency.

Initially the passive core-shell nanoresonators without dye molecules were studied. Various parameters were determined, including the average and maximum local **E**-field, the optical (absorption (*ACS*), scattering (*SCS*) and extinction (*ECS*)) cross-sections the gold absorption, and the power outflow (outflow) as a function of wavelength. Additionally, the full width at half maximum (*FWHM*) of these quantities was also calculated to evaluate the degree of spectral line narrowing in the active systems (Fig. S2).

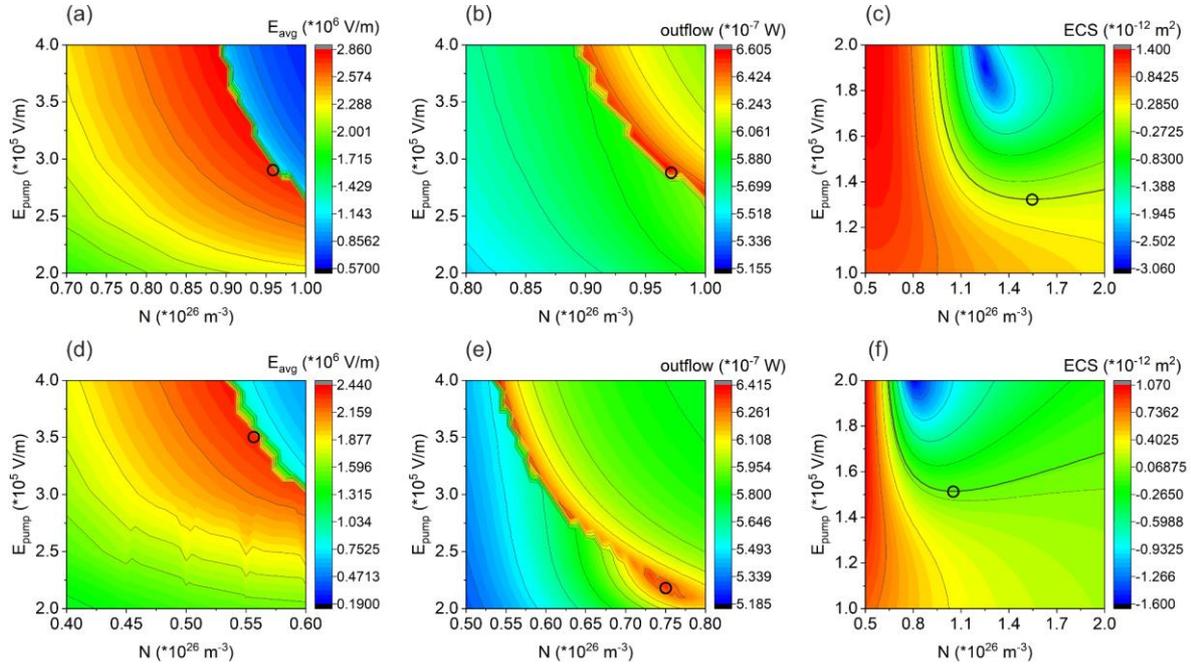

**Figure 1.** Objective functions over the pump **E**-field strength and Rh800 dye concentration parameter space. The pump **E**-field strength and concentration dependence of the (a, d) average local **E**-field, (b, e) power outflow and (c, f) extinction cross-section in the (a-c) GMD and (d-f) GMG composition. The black circles indicate the parameter-pairs with potential to achieve optimal characteristic in different operation regions, the contour lines are shown to guide the eyes. The thicker black contour on the map of the *ECS* indicates the *ECS*=0 condition.

In the case of the active systems, three different objective functions were used: (i) to maximize the near-field enhancement (NF-type); (ii) to maximize the power-outflow into the far-field (FF-type); and (iii) to minimize the spaser threshold in the pump **E**-field strength that corresponds to the extinction cross-section sign-switching (ECST-type). For reference purposes, analytical calculations were conducted considering the GAR quantifying the geometry and the composition of the GMD nanoresonator to determine the spaser threshold in the permittivity [Eq. (2) in Ref. 18 & Eq. (S3) in the Supplementary], as well as the minimal spaser threshold in the dye concentration that results in zero-crossing in *ECS* at the emission wavelength in the GMG composition [Eq. (11) in Ref. 17 & Eq. (S4) in the Supplementary]. To uncover their capabilities, the complete pump **E**-field ($E_{pump}$) and dye concentration (*N*) dependence of the average local **E**-field, the power-outflow and the *ECS* at the probe wavelength were mapped (Fig. 1).

Different regions were identified on the parameter plane, where the above described objective functions attain their maximum values. The selection criteria included that only a local **E**-field below the damage threshold of the nanoresonator is allowed [30].

The 2D maps of GMD and GMG compositions are analogous, only quantitative differences are observable. Based on the maps of the two inspected core-shell compositions similar average local **E**-field can be achieved by simultaneously modulating the pump **E**-field strength and dye concentration (contours in Fig. 1a, d). By increasing (decreasing) the concentration smaller (larger) pump **E**-field strength is required to achieve the same local **E**-field. Accordingly, the largest average local **E**-field can be achieved along a hyperbola in the parameter-plane. Below this hyperbola, the local **E**-field gradually decreases, whereas above it rapidly drops. On the map of the outflow a reversed trade-off can be seen between the pump **E**-field strength and the dye concentration (contours in Fig. 1b, e). The outflow rapidly and significantly decreases below the hyperbola of the maximal value, while above it the outflow gradually and slightly decreases. A trade-off can be recognized also on the map of the *ECS*, but the tendencies are fundamentally different (contours in Fig. 1c, f). The *ECS* is positive below a parabola-like *ECS*=0 contour-line, while above it takes on negative values. In the region of small concentrations, the large and positive value of the *ECS* is almost independent of the **E**-field strength of the pumping; at intermediate concentrations the rapidly varying negative *ECS* takes on a global minimum, while at larger concentrations the *ECS* monotonously decreases by increasing the **E**-field pump strength.

By selecting the appropriate $E_{pump}$-$N$ parameters, the average and maximal local **E**-field and near-field enhancement (*NFE*) inside the gain medium, the absorptions inside the gold shell and gain media, the total absorption, the outflow, the internal (*IQE*) and external (*EQE*) quantum efficiency, the *ACS* of the gold shell and the gain medium, the total *ACS*, the *SCS* as well as the *ECS* were inspected at the lasing wavelength. The spectral dependence and the *FWHM* of the extrema in these quantities were determined to calculate the degree of narrowing in active nanoresonators compared to their passive counterparts. For each objective functions, comparisons were realized between the two inspected compositions to determine the preferred nanoresonator types for different operation regions (Fig. S3).

## 3. Results and discussion

### 3.1 Wavelength dependent dielectric properties

The achieved population inversion rate has a considerable impact on the dielectric properties of every inspected nanoresonators and compositions, both at the absorption and emission wavelengths, except the real part of the permittivity at the pump wavelength, which remains unchanged (Table S2).

At the pump wavelength the imaginary part of the permittivity and the refractive index are positive, in accordance with the expected pump beam absorption in the dye -doped medium. The real part of the refractive index slightly increases compared to the host medium ($n_{host}$=1.62) (Table S2). The deviations from the host medium are noticeably larger in the GMD nanoresonators, except the FF-type NR, where the deviation in the (imaginary) real part of the refractive index is (almost) the same for the two nanoresonator compositions.

At the probe wavelength the spatially averaged imaginary parts become negative, which indicates the gain of the probe beam. The real part of the spatial average of the permittivity and refractive index slightly increases compared to the host medium (Table S2, Fig. 2-4a, b). The deviations from the host medium are larger in case of the GMD nanoresonators, except the FF-type NR, where the deviation in the real part of the permittivity and refractive index is larger in the GMG composition.

### 3.2 Local E-field and near-field enhancement

The pump **E**-field strength was $2.9*10^5$ and $3.5*10^5$ / $2.9*10^5$ and $2.2*10^5$ / $1.32*10^5$ and $1.52*10^5$ V/m for the GMD and GMG nanoresonators in NF-type / FF-type / ECST-type NRs respectively.

This resulted in a 17.2-fold and 13.1-fold / 17.1-fold and 20.9-fold / 11.1-fold and 8.0-fold average pump enhancement in the dye media. The enhancement is larger for the GMD composition, except the FF-type NRs, where the enhancement is slightly larger for the GMG nanoresonator (Table S3).

In the local **E**-field spectra a single peak appears at the lasing wavelength in the NF-type NRs (Fig. 2c), while the peaks slightly blue-shifts in the FF-type NRs (Fig. 3c) and more pronouncedly blue-shift in the ECST-type NRs (Fig. 4c). In all cases an additional shoulder is observable near 720 nm.

Significantly larger near-field enhancement can be achieved at the probe beam wavelength. The average (maximum) *NFE* is 276 (281) -fold and 234 (392) -fold / 116 (118) -fold and 147 (240) -fold / 85.4 (87.1) -fold and 60.1 (93.7) -fold in case of GMD and GMG nanoresonators in NF-type / FF-type / ECST-type nanoresonators respectively. The average (maximal) *NFE* is larger in the GMD (GMG) composition in NF-type and ECST-type NRs, while both quantities are larger for the GMG composition in FF-type nanoresonators. The *NFE* is more uniform in the GMD nanoresonators, based on the more comparable average and maximal **E**-field values in all NR types (Table S3, Fig. 1a, d).

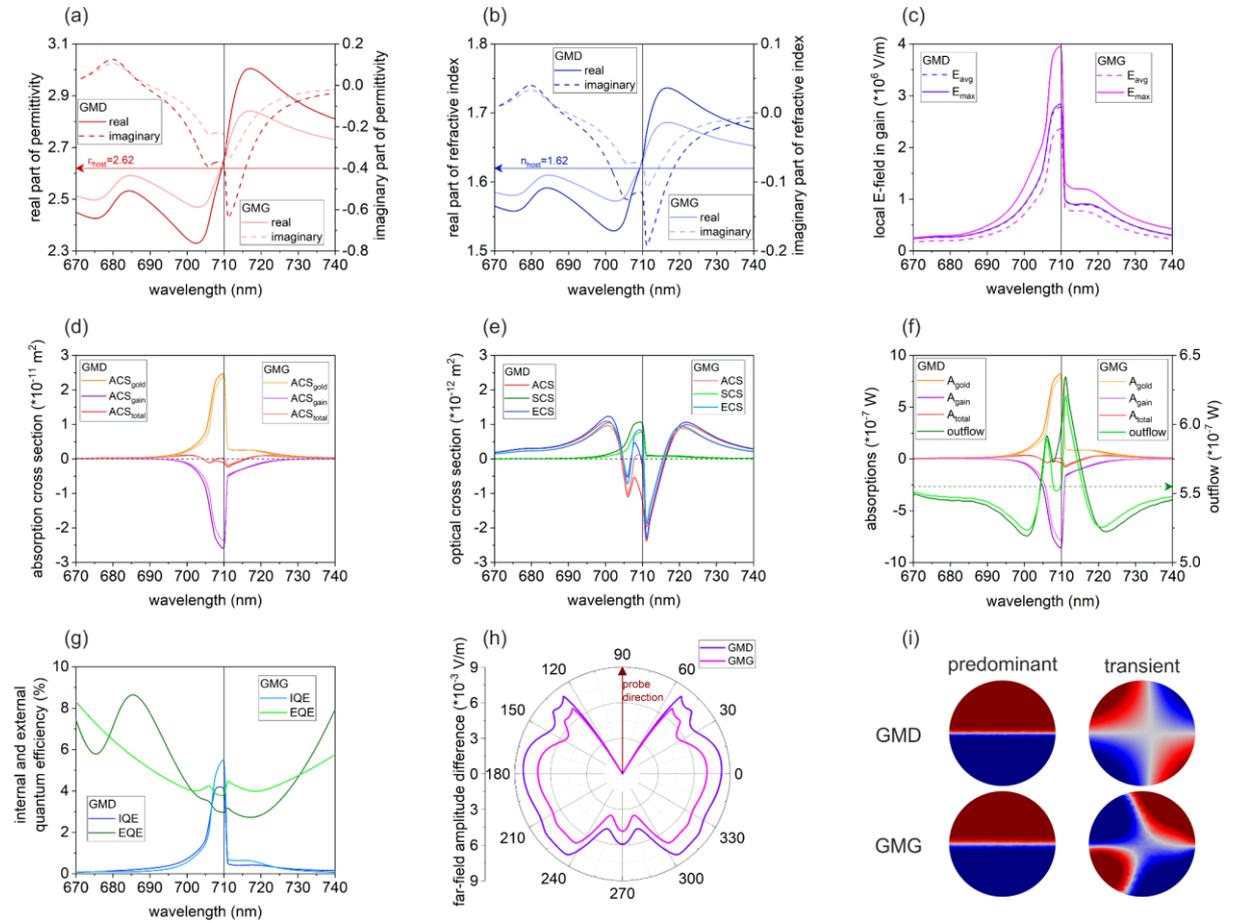

**Figure 2.** Spectra, far-field and surface charge distribution of the NF-type nanoresonators. The spectral dependence of the (a) permittivity and (b) refractive index of the dye seeded dielectric media, (c) average ($E_{avg}$) and maximal ($E_{max}$) local **E**-field, (d) absorption cross-sections, (e) complete optical cross-sections, (f) gold, gain and total absorption and outflow, and (g) internal and external quantum efficiency. The (h) polar angle distribution of the far-field amplitude, and the (i) predominant and transient surface charge distribution at the lasing wavelength.

*FWHM* decrease is achieved in every inspected nanoresonators compared to the passive counterparts in all **E**-field spectra (Fig. S2a). The *FWHM* is slightly larger (the same) / smaller (smaller) / smaller (larger) in the average (maximal) local **E**-field spectrum of the GMD composition compared to the **E**-field spectrum of the GMG in NF-type / FF-type / ECST-type nanoresonators (Table S6).

Larger (smaller) narrowing is achieved with respect to the passive counterparts via the GMD composition in the average (maximal) local **E**-field spectra compared to the GMG composition in all inspected nanoresonator types (Table S7).

### 3.3 Optical cross sections

Both in the gold and gain *ACS* spectra narrow resonances appear, that manifest themselves as a peak and dip around the lasing wavelength in the NF-type NR (Fig. 2d), while the extrema are slightly and more pronouncedly blue-shifted in the FF-type (Fig. 3d) and ECST-type nanoresonators, respectively (Fig. 4d). In all cases an additional shoulder is observable near 720 nm in the gold *ACS* spectra. The absolute values of the gain *ACS* are larger than that of the gold *ACS* in a wide spectral interval, which results in negative total *ACS* at the emission wavelength, except the GMG composition in the NF-type NRs. Although, in the GMG-NF nanoresonator the gold *ACS* is slightly larger than the absolute value of the gain *ACS*, which results in a small, but still positive total *ACS* at the emission wavelength, the spectra show that the GMG-NF also act as a near-field amplifier in proximity of 710 nm.

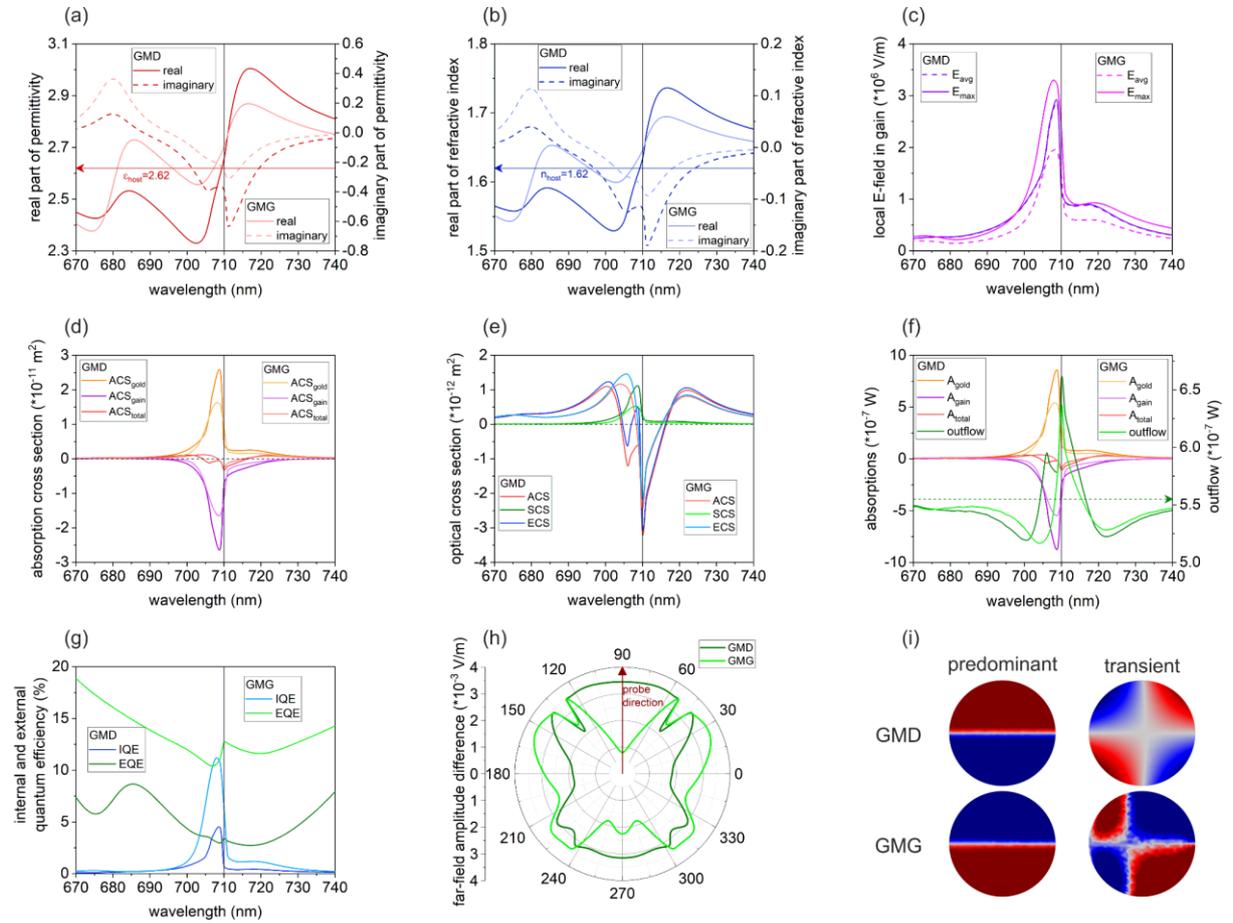

**Figure 3.** Spectra, far-field and surface charge distribution of the FF-type nanoresonators. The spectral dependence of the (a) permittivity and (b) refractive index of the dye seeded dielectric media, (c) average ($E_{avg}$) and maximal ($E_{max}$) local **E**-field, (d) absorption cross-sections, (e) complete optical cross-sections, (f) gold, gain and total absorption and outflow, and (g) internal and external quantum efficiency. The (h) polar angle distribution of the far-field amplitude, and the (i) predominant and transient surface charge distribution at the lasing wavelength.

In both compositions of the NF-type NRs the total *ACS* spectra are split, resulting in a local maximum near 710 nm, and a local minimum below and a global minimum above 710 nm (Fig. 2f). In FF-type GMD nanoresonators the total *ACS* spectrum is similarly split, leading to a local minimum below and a global minimum at 710 nm, whereas a single global minimum appears at the lasing wavelength in the total *ACS* of the GMG composition (Fig. 3f). In the ECST-type NRs no splitting occurs in the total *ACS* spectra, only a single small global minimum appears at 710 nm (Fig. 4f).

In the *SCS* spectra of the NF-type nanoresonators a single peak appears at the emission wavelength (Fig. 2f). Conversely, the *SCS* peaks are slightly and more pronouncedly blue-shifted in the FF-type (Fig. 3f) and ECST-type (Fig. 4f) NRs. In the NF-type nanoresonators the *SCS* is smaller than the absolute value of the total *ACS* in the GMD composition, while it is significantly larger in the GMG nanoresonator. The *SCS* of the FF-type NRs is significantly smaller than the absolute value of the total *ACS*, while in the ECST-type nanoresonators the value of the *SCS* equals the total *ACS* for both NR compositions (Table S4).

The *ECS* is governed mainly by the *ACS*, accordingly their spectral dependencies are similar. Namely, in the NF-type NRs a local maximum appears near 710 nm and a local and global minimum is observable below and above it (Fig. 2f). In the FF-type nanoresonators the *ECS* spectra are qualitatively different, as a local and global minimum appears below and at 710 nm in the *ECS* of the GMD, whereas a global minimum develops at 710 nm in the *ECS* of the GMG (Fig. 3f). In either of the ECST-type NRs there is no splitting around the emission wavelength in the *ECS* spectra, only a single small global minimum appears at 710 nm, similarly to the total *ACS* (Fig. 4f).

In the NF-type GMD nanoresonator at the lasing wavelength negative *ECS* is achieved due to the negative total *ACS*, while the *ECS* remains positive in the GMG due to the small positive total *ACS*. In the FF-type NRs the *ECS* is negative for both compositions at the lasing wavelength, indicating that both have already entered into the spasing operation region. In the ECST-type nanoresonators the *ECS* is approximately zero at 710 nm due to that the values of the *SCS* and the total *ACS* are equal. The *ECS* becomes negative in the GMD nanoresonator slightly above the lasing wavelength, while it remains positive nearby 710 nm in the GMG. As it will be detailed in succeeding paragraphs, on the average the GMD nanoresonator exhibits better spaser characteristic, though GMG better approximates the zero-crossing corresponding to the criterion of spasing region. By comparing the spaser thresholds in terms of the pump **E**-field strength and dye concentration, the GMD nanoresonator exhibits the zero-crossing at a smaller pump **E**-field strength ($1.32*10^5$ V/m < $1.52*10^5$ V/m in Table S3), but at a larger dye concentration ($1.535*10^{26}$ m$^{-3}$ > $1.05*10^{26}$ m$^{-3}$ in Table S2).

According to the analytical calculations, considering both the geometry and the composition [18], the gain threshold is -0.393 for the inspected GMD nanoresonator, which is close to the numerically derived threshold of the ECST nanoresonator ($\varepsilon_{imag\_GMD}(\lambda_e)$=-0.374 in Table S2 and Fig 4a). The minimal gain threshold for spasing calculated based on the dielectric properties of composing materials is -0.165 at 710 nm [17], which is smaller than the threshold of the inspected GMG NR ($\varepsilon_{imag\_GMG}(\lambda_e)$=-0.188 in Table S2 and Fig. 4a). The minimal gain threshold could be achieved at smaller dye concentration but at a larger pump **E**-field strength (Fig. 1c, f).

Larger gain and gold *ACS* as well as *SCS* are achieved with the GMD composition in the NF-type and ECST-type NRs, while these quantities are smaller in the FF-type GMD nanoresonator. The absolute value of the total *ACS* and the *ECS* are larger in the GMD composition, except the NF-type nanoresonators, where the *ECS* is larger in the GMG composition (Table S4).

*FWHM* narrowing is achieved in every inspected nanoresonators compared to their passive counterparts (Fig. S2b) which is an indicator of a lasing medium. The *FWHM* of the gold as well gain *ACS* extrema is larger / smaller / smaller in NF-type / FF-type / ECST-type GMD nanoresonators, respectively according to the achieved smaller / larger / larger line-width narrowing in the gold *ACS*. The *FWHM* all of the total *ACS*, *SCS* and *ECS* extrema is larger in the GMD nanoresonator, according to the smaller narrowing compared to the passive counterparts in NF-type NRs. The *FWHM* is larger / smaller of the extrema appearing in the total *ACS* and *ECS*, and smaller / larger of the peak reached in the *SCS*, as a result of the achieved smaller / larger narrowing in the total *ACS* and *ECS* extrema, and larger / smaller narrowing in the *SCS* peak in FF-type / ECST-type GMD nanoresonators, respectively (Table S6, S7).

### 3.4 Study on the optical responses, quantum efficiencies, far-field and charge distribution

In case of the NF-type nanoresonators in the gold and the gain absorption spectra a narrower peak and a dip appears at the emission wavelength (Fig. 2f). Conversely, the peaks and dips are slightly and more pronouncedly blue-shifted in the FF-type (Fig. 3f) and ECST-type (Fig. 4f) nanoresonators, respectively. In all cases an additional shoulder is observable near 720 nm in the gold absorption spectra in accordance with the local **E**-field enhancement. The gain is larger than the gold absorption, resulting in negative total absorption at 710 nm, except the GMG composition of the NF-type NRs. Though, the gold absorption remains slightly larger than the gain, leading to a tiny, but positive total absorption at the emission wavelength, the zero-crossing occurring at a slightly redshifted spectral location proves loss-compensating operation also in the NF-type GMG.

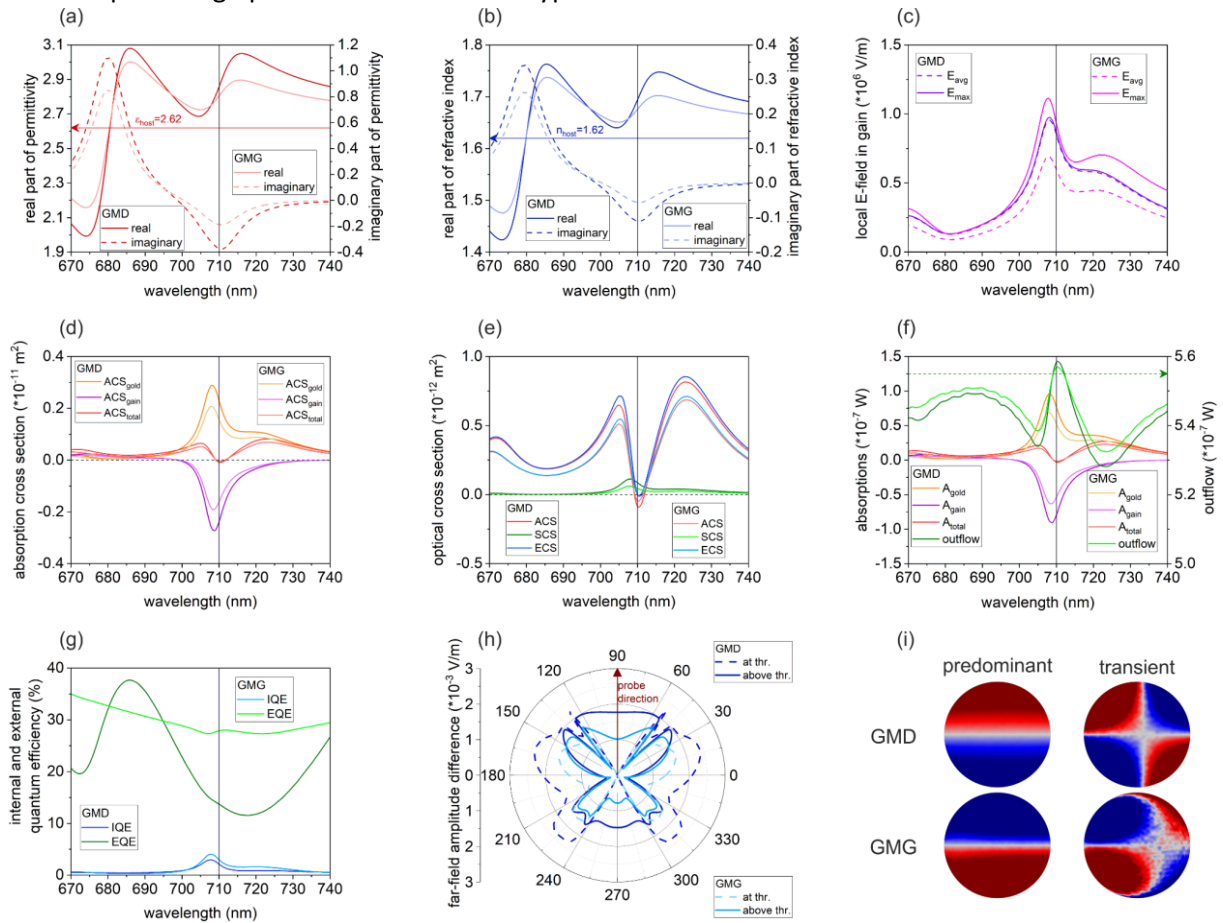

**Figure 4.** Spectra, far-field and surface charge distribution of the ECST-type nanoresonators. The spectral dependence of the (a) permittivity and (b) refractive index of the dye seeded dielectric media, (c) average ($E_{avg}$) and maximal ($E_{max}$) local **E**-field, (d) absorption cross-sections, (e) complete optical cross-sections, (f) gold, gain and total absorption and outflow, and (g) internal and external quantum efficiency. The (h) polar angle distribution of the far-field amplitude, and the (i) predominant and transient surface charge distribution at the lasing wavelength.

In the NF-type nanoresonators the total absorption spectrum is split for both compositions, which manifests itself in a local maximum near 710 nm, and a local and global minimum below and above 710 nm, respectively (Fig. 2f). Conversely, in the FF-type NRs the total absorption spectrum is qualitatively different (Fig. 3f). The absorption spectrum is split only in the GMD nanoresonator, which results in a local and global minimum below and at 710 nm, respectively. In contrast, a single shallow global minimum appears at the lasing wavelength in the total absorption of the GMG nanoresonator.

In the ECST-type nanoresonators no splitting occurs in the total absorption spectra, only a single small global minimum appears at 710 nm that is relatively deeper than in the FF-type GMG nanoresonator (Fig. 4f).

Due to the negative total absorption the outflow is enhanced compared to the incoming probe intensity in a considerable spectral interval around the emission wavelength. Exception is the GMG composition of the NF-type nanoresonators, where the outflow is slightly decreased in close proximity of 710 nm, due to the very small, but still positive total absorption at the emission wavelength.

The spectral dependence of the outflow is governed by that of the total absorption spectra (Fig. 2-4f). As a result, in the NF-type nanoresonators a local minimum appears slightly below 710 nm, between a local and global maximum developing well-below and slightly above 710 nm, respectively. In the FF-type nanoresonators the spectra qualitatively differ, as a local and global maximum appears below and at 710 nm in the GMD NR, respectively, whereas a non-split global maximum is observable at 710 nm in the GMG NR. In the ECST-type nanoresonators the spectra are more likely to that of the FF-type GMG NR, namely a non-split global maximum appears in the outflow exactly at 710 nm.

Slightly larger gain and gold absorption are achieved with the GMD nanoresonators in the NF-type and ECST-type nanoresonators, while both quantities are significantly smaller in GMD composition of the FF-type nanoresonators. The absolute value of the total absorption and the outflow are consistently larger in the GMD composition (Table S5).

*FWHM* narrowing is achieved in every inspected nanoresonators compared to their passive counterparts (Fig S2c). For the GMD composition the *FWHM* of the peak in the gold absorption and dip in the gain absorption is larger / smaller / slightly smaller, and the *FWHM* of the dip in the total absorption and peak in the outflow is larger / larger / smaller compared to the GMG composition in NF-type / FF-type / ECST-type nanoresonators, respectively (Table S6). Accordingly, the achieved narrowing is smaller / larger / slightly larger in the gold absorption peaks and smaller / smaller / larger in the total absorption dips and outflow peaks, compared to the passive counterpart nanoresonators (Table S7).

At the lasing wavelength the nanorasonators optimized to maximize the NFE exhibit a global *IQE* maximum and a local *EQE* minimum (Fig. 2g), in accordance with their near-field confinement capability and demonstrating their potential to act as near-field amplifiers. Conversely, the nanoresonators optimized to maximize the power-outflow exhibit a decreasing *IQE* tendency at the lasing wavelength, as the *IQE* shows a global maximum at a slightly smaller wavelength, while the *EQE* shows a local maximum (Fig. 3g). According to the far-field emission of the highest intensity, these systems can act as nanolasers with considerably improved out-coupling efficiency. The nanoresonators optimized to minimize spaser threshold exhibit a decreasing *IQE* tendency, as the *IQE* shows a global maximum at a slightly smaller wavelength, and an *EQE* value succeeding a global maximum / preceding local maximum in the spectra of GMD / GMG composition, respectively (Fig. 4g). Accordingly, these structures can act as spasers with a slightly different characteristics. In the NF-type nanoresonators the *IQE* exceeds the *EQE* at the emission wavelength for both compositions according to the near-field operation, while in the FF-type and ECST-type NRs the *EQE* outperforms the *IQE* – indicating the improved potential for light extraction. Both the *IQE* and *EQE* are smaller for the GMD compared to the GMG composition, indicating that the GMG composition promotes slightly better both the near-field confinement and the out-coupling efficiencies, even though the achieved outflow is larger for the GMD composition in all inspected nanoresonator types (Table S5).

The far-field distributions are similar in the two inspected nanoresonator compositions for the NF-type and ECST-type nanoresonators, whereas the distributions qualitatively differ in the case of the FF-type NRs. The nanoresonators optimized to maximize near-field enhancement amplify the radiation almost uniformly, except in a 60° interval around the probe direction (Fig. 2h). In contrast, the nanoresonators optimized to maximize the outflow amplify the radiation in the complete polar angle region, including the probe direction (Fig. 3h). Remarkable difference is that the GMG allows for a tiny forward radiation, whereas in the GMD the highest emission is achieved along the forward scattering direction.

The nanoresonators optimized to minimize the spaser threshold amplify the radiation in several discrete polar angle intervals, however the emission maxima are oriented almost perpendicularly to the probe beam direction (Fig. 4h). By increasing the pump **E**-field strength, considerable radiation enhancement can be achieved also in the probe beam direction. The GMD nanoresonator produces larger far-field lobes in all inspected nanoresonator types.

The radiation pattern characteristic is similar in NF-type and ECST-type nanoresonators, whereas it differs significantly in the FF-nanoresonators. Namely, the far-field radiation becomes maximal around 5° / 20° azimuthal orientation for both the GMD and GMG composition in NF-type / ECST-type nanoresonators, while it is maximal along the probe beam propagation direction (at 30°) in the FF-type GMD (GMG) nanoresonators. The radiation pattern characteristic slightly modifies by increasing the pump **E**-field strength in the ECST-type NRs, namely the radiation becomes maximal around 55° (40°) in GMD (GMG) composition.

The charge distribution is dominantly dipolar at the lasing wavelength in all inspected NRs. However, when examining the time-evolution of the charge distribution, it becomes evident that quadrupolar separation also appears intermittently during each duty cycles (Fig. 2-4i). By increasing the concentration and decreasing the pump **E**-field strength, the fraction of quadrupolar to dipolar distribution (*QP:DP* ratio) can be preserved, but the larger QP fraction is advantageous to enter into the spaser operation region, as higher order modes are excitable at lower threshold (Fig. 1). By comparing the two nanoresonator compositions, the GMD nanoresonators produced larger *QP:DP* ratio compared to their GMG counterparts both in the passive and the active configurations, and for all three inspected nanoresonator types (Table S5).

## 4. Conclusions

A comparative study on the two different active core-shell nanoresonators, namely the GMD and GMG compositions, was realized. By mapping the pump **E**-field strength and the dye molecule concentration dependence in a pre-optimized geometry, distinct regions of particular interest can be identified. These regions correspond to the maximal near-field enhancement (NF-type), to the highest power-outflow (FF-type), or to the minimal spaser threshold in the pump **E**-field strength (ECST-type), that can be reached with the criterion that the local **E**-field is below the damage threshold of the nanoresonators (Fig. 1).

Based on the results, the NF-type GMD is better in the improvement of quantities particularly important to NF-type NRs (average **E**-field enhancement and line-width narrowing). In addition to it the NF-type GMD exhibits better improved FF-type specific quantities (larger gain and total *ACS* & absorption, outflow and *SCS*) and possesses better spaser threshold behavior relevant quantities (smaller negative *ECS* and pump **E**-field strength) as well. In comparison, the GMG is better in all other (two-times larger amount of) quantities, including NF-type relevant quantities (maximal **E**-field enhancement and narrowing), FF-type relevant quantities (smaller $ACS_{gold}$ and $A_{gold}$, as well as larger *IQE* and *EQE*), and all *FWHM* and narrowing improvements (except the *FWHM* of the maximal **E**-field and narrowing in the average **E**-field). These properties reveal that the GMG allows for near-field amplifier operation with smaller loss and better promotes line-width narrowing.

The FF-type GMD and FF-type GMG compositions are more comparable, e.g. the FF-type specific power-outflow is only slightly larger in GMD (in addition to the smaller gold but larger total *ACS* and absorption, smaller *FWHM* in $ACS_{gold}$ & $A_{gold}$, $ACS_{gain}$ & $A_{gain}$ and *SCS* due to the more pronounced narrowing). The FF-type GMD exhibits NF-type relevant advantages (smaller *FWHM* accompanied by larger narrowing in the average **E**-field and smaller *FWHM* in the maximal **E**-field) as well as it is considerable in post-threshold spaser operation (larger absolute value of *ECS*). In comparison, the GMG is endowed with NF-type relevant advantages (larger *IQE*, average and maximal **E**-field enhancement and narrowing in maximal **E**-field line*)*, results in better FF-type relevant quantities (larger *ACS$_{gain}$* & *A$_{gain}$*, *SCS*, *EQE*, smaller *FWHM* in *ACS$_{total}$* and *ECS*, as well as in *A$_{total}$* and outflow due to the more pronounced narrowing). It also possesses advantages for spaser threshold behavior (smaller *E$_{pump}$* and smaller absolute value of *ECS*). These properties reveal that the GMG allows for larger out-coupling efficiency and is endowed with good near-field amplifier and low-threshold spaser properties.

The threshold of **E**-field pump strength that is particularly important for ECST-type nanoresonators is smaller in GMD composition. In addition the ECST-type GMD is slightly better in the improvement of NF-type relevant quantities (average **E**-field enhancement, *FWHM* and narrowing), and is better in FF-type relevant quantities (larger gain and total *ACS* and absorption, *SCS* and outflow, smaller *FWHM* as well as larger narrowing in the *OCS* and optical responses – except the *SCS*) as well. In comparison, the ECST-type GMG is advantageous in NF-type relevant properties (better maximal **E**-field with smaller FWHM due to better narrowing, larger *IQE*), and in FF-relevant properties (significantly larger *EQE*, smaller *SCS FWHM* due to better narrowing, smaller *A$_{gold}$* and *ACS$_{gold}$*). Although, its *E$_{pump}$* threshold is larger the achieved *ECS* better approaches zero in GMG, which is an important property for small-threshold spaser operation. These properties reveal that the GMG allows for high internal and external efficiency.

In conclusion, based on the objective functions, the GMD composition produced larger average near-field enhancement in the NF-type and higher power-outflow in the FF-type nanoresonators, and also exhibited a smaller spaser threshold in the pump **E**-field strength as an ECST-type NR – but resulted in a slightly larger minimum ECS as well. Taking into account the other inspected quantities the GMG composition can be better in NF-type and FF-type NRs and is competitive as ECST-type NRs as well. This is due to the composition with smaller amount of gold and larger amount of gain. Both nanoresonator compositions are suitable for near-field amplifier, nanolasing and spasing operation purposes.

**Author Contributions**

**Dávid Vass:** Data curation, Formal analysis, Investigation, Methodology, Visualization, Writing – original draft **András Szenes:** Data curation, Investigation, Methodology **Nagy Péter Zoltán:** Data curation, Investigation **Balázs Bánhelyi:** Software **Mária Csete**: Conceptualization, Supervision, Investigation, Writing – review & editing

**Declaration of competing interest**

The authors declare that they have no known competing financial interests or personal relationships that could have appeared to influence the work reported in this paper

**Data availability**

Data will be made available on request.

**Funding**

This work was supported by the National Research, Development and Innovation Office (NKFIH) of Hungary, projects: "Optimized nanoplasmonics" (K116362), "Nanoplasmonic Laser Inertial Fusion Research Laboratory" (NKFIH-2022-2.1.1-NL-2022-00002) and "National Laboratory for Cooperative Technologies" (NKFIH-2022-2.1.1-NL-2022-00012) in the framework of the Hungarian National Laboratory program.

**References**

[1] E. Prodan at. al.: "A Hybridization Model for the Plasmon Response of Complex Nanostructures", Science 302, 419 (2003). https://doi.org/10.1126/science.1089171

[2] N. Halas: "Playing with Plasmons: Tuning the Optical Resonant Properties of Metallic Nanoshells" Mrs. Bulletin 30 (2005). https://doi.org/10.1557/mrs2005.99

[3] R. Bardhan et. al.: "Metallic Nanoshells with Semiconductor Cores: Optical Characteristics Modified by Core Medium Properties" ACS Nano 4, 6169-6179 (2010). https://doi.org/10.1021/nn102035q

[4] Y. Chen et. al.: "The Study of Surface Plasmon in Au/Ag Core/Shell Compound Nanoparticles", Plasmonics 7, 509-513 (2012). https://doi.org/10.1007/s11468-012-9336-6

[5] F. Tam et al.: "Mesoscopic nanoshells: Geometry-dependent plasmon resonances beyond the quasistatic limit", J. Chem. Phys. 127, 204703 (2007). https://doi.org/10.1063/1.2796169

[6] J. A. Schuller et al.: "Plasmonics for extreme light concentration and manipulation", Nat. Mater. 9 (2010). https://doi.org/10.1038/nmat2630

[7] A. Szenes. et. al.: „Improved emission of SiV diamond color centers embedded into concave plasmonic core-shell nanoresonators", Sci. Rep. 7, 13846 (2017). https://doi.org/10.1038/s41598-017-14227-w

[8] S. I. Azzam et. al.: "Ten years of spasers and plasmonic nanolasers ", Light Sci. Appl. 9, 90 (2020). https://doi.org/10.1038/s41377-020-0319-7

[9] C. Fietz et, al.: "Finite element simulation of microphotonic lasing system", Opt. Express 20, 11548 (2012). https://doi.org/10.1364/OE.20.011548

[10] R. Marani et. al.: "Gain-assisted extraordinary optical transmission through periodic arrays of subwavelength apertures", New Journal of Physics 14, 013020 (2012). https://doi.org/10.1088/1367-2630/14/1/013020

[11] M. Kim et. al.: "Frequency-domain modelling of gain in pump-probe experiment by an inhomogeneous medium", J. Phys. Condens. Mater. 30, 064003 (2017). https://doi.org/10.1088/1361-648X/aaa473

[12] S.-L. Wang et. al.: "Loss and gain in a plasmonic nanolaser", Nanophotonics 9, 3403-3408 (2020). https://doi.org/10.1515/nanoph-2020-0117

[13] R. Wang et. al.: "Plasmon–exciton coupling dynamics and plasmonic lasing in a core–shell nanocavity", Nanoscale 13, 6780 (2021). https://doi.org/10.1039/D0NR08969A

[14] Y. Li et. al.: "Low Threshold and Long-Range Propagation Plasmonic Nanolaser Enhanced by Black Phosphorus Nanosheets", Adv. Theory Simul. 4, 2100087 (2021). https://doi.org/10.1002/adts.202100087

[15] M. A. Noginov et. al.: "Demonstration of a spaser-based nanolaser", Nature Letters 460, 1110-1112 (2009). https://doi.org/10.1038/nature08318

[16] Y. Huo et. al.: "Spaser based on dark quadrupole surface plasmon mode of a trapezoidal nanoring", Optics Commun. 465, 125485 (2020). https://doi.org/10.1016/j.optcom.2020.125485

[17] N. Arnold et. al.: "Minimal spaser threshold within electrodynamic framework: Shape, size and modes", Ann. Phys. 528, 2965-306 (2016). https://doi.org/10.1364/OME.5.002546

[18] N. Arnold et. al.: "Spasers with retardation and gain saturation: electrodynamic description of fields and optical cross-sections", Opt. Mater. Express 5, 2546-2577 (2015). https://doi.org/10.1364/OME.5.002546

[19] G. V. Kristanz et. al.: "Power balance and temperature in optically pumped spasers and nanolasers", ACS Photonics 5, 3696-3703 (2018). https://doi.org/10.1021/acsphotonics.8b00705

[20] A. Szenes et. al.: "Active Individual Nanoresonators Optimized for Lasing and Spasing Operation", Nanomaterials 11, 1322 (2021). https://doi.org/10.3390/nano11051322

[21] V. Caligiuri et. al.: "Resonant Gain Singularities in 1D and 3D Metal/Dielectric Multilayered Nanostructures", ACS Nano 11, 1012-1025 (2017). https://doi.org/10.1021/acsnano.6b07638

[22] S. Ning et. al.: "Low threshold and wavelength tunable SPASER based on plasmonic coupled nanostructure", Journal of Luminescece 252, 119286 (2022). https://doi.org/10.1016/j.jlumin.2022.119286

[23] J.-S. Wu et. al.: "Topological Spaser", Phsy. Rev. Lett. 124, 017701 (2020). https://doi.org/10.1103/PhysRevLett.124.017701

[24] S. Lee et. al.: "Random Lasing with a High Degree of Circular Dichroism by Chiral Plasmonic Gold Nanoparticles", ACS Photonics 9, 613-620 (2022). https://doi.org/10.1021/acsphotonics.1c01601

[25] S. F. Haddawi et. al.: "Coupled modes enhance random lasing in plasmonic double grating structure", Optics & Laser Technology 156, 108577 (2022). https://doi.org/10.1016/j.optlastec.2022.108577

[26] Z. Zhang et. al.: "Lasing-enhanced surface plasmon resonance spectroscopy and sensing", Photonics Research 9, 1699-1714 (2021). https://doi.org/10.1364/PRJ.431612

[27] N. M. Ngo et. al.: "Plasmonic Nanostars: Systematic Review of their Synthesis and Applications", ACS Appl. Nano Mater. 5, 14051-14091 (2022). https://doi.org/10.1021/acsanm.2c02533

[28] D. Vass et. al.: "Plasmonic nanoresonator distributions for uniform energy deposition in active targets", Opt. Mater. Express 13, 9-27 (2023). https://doi.org/10.1364/OME.471980

[29] T. Csendes et. al.: "The GLOBAL optimization method revisited", Optim. Lett. 2, 445-454 (2008). https://doi.org/10.1007/s11590-007-0072-3

[30] A. M. Fales et. al.: "Quantitative Evaluation of Nanosecond Pulsed Laser-Induced Photomodification of Plasmonic Gold Nanoparticles", Sci. Rep. 7, 15704 (2017). https://doi.org/10.1038/s41598-017-16052-7

# Supporting Information

*Dávid Vass[1,2], András Szenes[1,2], Péter Zoltán Nagy[1], Balázs Bánhelyi[2,3] and Mária Csete[1,2*]*

[1]*University of Szeged, Department of Optics and Quantum Electronics, Szeged 6720, Hungary*
[2]*Wigner Research Centre for Physics, Budapest 1121, Hungary*
[3]*University of Szeged, Department of Computational Optimization, Szeged 6720, Hungary*
*\*mcsete@physx.u-szeged.hu*

## 1. Passive optimized core-shell configurations

The geometries of the optimized core-shell compositions are slightly different (Table S1). The core radius is larger, while the dielectric coating thickness is smaller in the optimized gain-metal-dielectric (GMD) composition, while the metal shell thickness is the same. Accordingly, the generalized aspect ratio (*GAR*) is slightly larger in the GMD nanoresonator i.e. the gold volume is larger, while the gain volume is smaller in the GMD.

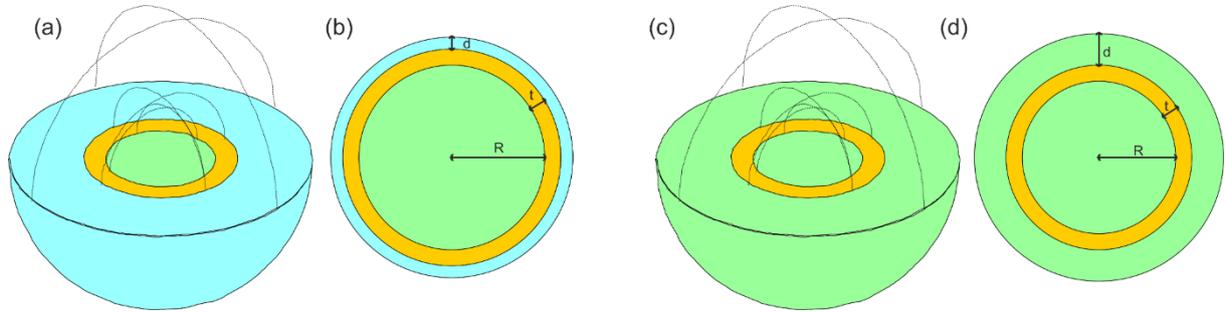

**Figure S1.** Schematic figures of the inspected compositions. The (a, c) 3D schematic figure of the core-shell nanoresonators, and the (b, d) xy plane cross-sections in the (a, b) gain-metal-dielectric and (c, d) gain-metal-gain composition.

| Geometry | | |
|---|---|---|
| | GMD | GMG |
| R (nm) | 28.7 | 23.7 |
| t (nm) | 5 | 5 |
| d (nm) | 3.7 | 9.8 |
| GAR | 0.85 | 0.83 |
| $V_{gold}$ (nm$^3$) | 6.1*10$^4$ | 4.3*10$^4$ |
| $V_{gain}$ (nm$^3$) | 9.9*10$^4$ | 2.0*10$^5$ |

**Table S1.** The geometrical parameters of the optimized compositions. The radius of the core (*R*), the gold shell thickness (*t*), the thickness of the outer dielectric coating (*d*), the generalized aspect ratio (*GAR*), as well as the gold ($V_{gold}$) and the gain ($V_{gain}$) volume.

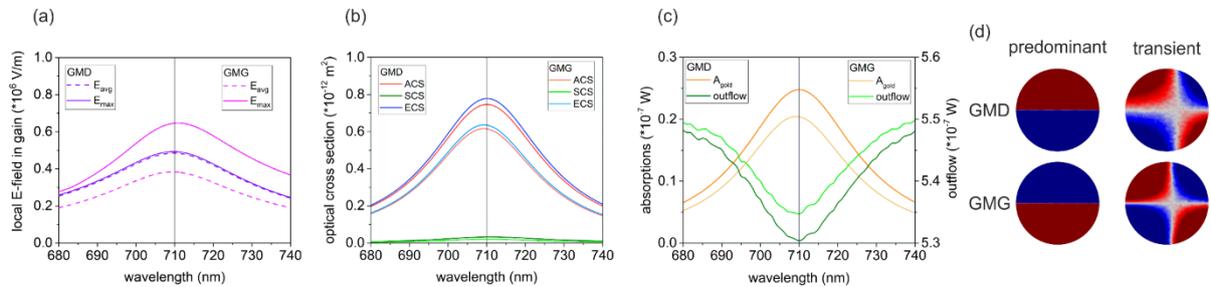

**Figure S2.** Spectra of the passive nanoresonators and the charge distributions. The spectral dependency of the (a) local average ($E_{avg}$) and maximal ($E_{max}$) **E**-field, the (b) optical cross-sections and the (c) optical response. (d) The predominant and transient charge distribution at the lasing wavelength.

In the local **E**-field spectra a single peak appears at the 710 nm emission wavelength. Larger average and smaller maximal local **E**-field is achieved in case of the GMD nanoresonator compared to the GMG composition. The *FWHM* is slightly larger (smaller) in the average (maximal) local **E**-field achieved via the GMD composition compared to the GMG (Fig. S2a).

In the *OCS* (*SCS*, *ACS* and *ECS*) spectra a single peak appears. Both NRs produce tiny *SCS* and large *ACS*. Due to this behavior, the *ECS* is determined by the *ACS*. All quantities are larger in GMD composition. The *FWHM* of these quantities is larger in the GMD compared to the GMG composition (Fig. S2b).

Similarly, a single peak appears in the absorption spectra. The absorption is larger in GMD nanoresonator. The outflow has a minimum at 710 nm. Due to the larger absorption the achieved outflow is smaller in case of the GMD. The *FWHM* is larger of the peaks both in the absorption and the outflow for the GMD composition compared to the GMG (Fig. S2c).

The charge distribution is dominantly dipolar at 710 nm, but the time evolution of the charge distribution shows that quadrupolar separation also appears intermittently during a duty cycle. By comparing the two compositions, the GMD nanoresonators produced larger *QP:DP* ratio compared to their GMG counterparts (Fig. S2d).

## 2. Analytical calculations for the spaser threshold

Based on Ref. [18], the gain threshold can be determined by analytical expressions for the GMD composition. The **E**-field inside the core can be expressed as a function of the external field and the dielectric and geometrical properties of the core-shell nanoresonator. From the denominator (*D*) of this expression, the threshold gain can be determined as follows:

$$D = (\varepsilon_1 + 2\varepsilon_2)(\varepsilon_2 + 2\varepsilon_3) + 2f(\varepsilon_1 - \varepsilon_2)(\varepsilon_2 - \varepsilon_3), \qquad (S1)$$

where $\varepsilon_1$ is the permittivity of the core medium, $\varepsilon_2$ is the permittivity of the shell medium, $\varepsilon_3$ is the permittivity of the host medium and *f* is the cubic ratio of the inner and outer radii:

$$f = \frac{r_1^3}{r_2^3} = \frac{R^3}{(R+t)^3}. \qquad (S2)$$

From these equations in case of *D*=0 the threshold gain can be calculated as follows:

$$\varepsilon_{thr\_GMD} = -i\left(\varepsilon_h + 2\varepsilon_2 \frac{(\varepsilon_2 + 2\varepsilon_3) - f(\varepsilon_2\varepsilon_3)}{(\varepsilon_2 + 2\varepsilon_3) + 2f(\varepsilon_2 - \varepsilon_3)}\right). \qquad (S3)$$

Based on Ref. [17] the gain threshold can be derived by analytical expressions also for the GMG composition from the polarizability. In this case the threshold gain can be written as:

$$\varepsilon_{thr\_GMG} = \frac{\varepsilon_{imag\_M}}{\varepsilon_{real\_M}} \varepsilon_{real\_G}, \qquad (S4)$$

where $\varepsilon_{real\_M}$ and $\varepsilon_{imag\_M}$ is the real and imaginary part of the metal shell permittivity, and $\varepsilon_{real\_G}$ is the real part of the gain permittivity.

| Dye concentration, population differences and dielectric properties | | | | | | |
|---|---|---|---|---|---|---|
| | NF-type | | FF-type | | ECST-type | |
| | GMD | GMG | GMD | GMG | GMD | GMG |
| N (*$10^{26}$ m$^{-3}$) | 0.96 | 0.56 | 0.97 | 0.75 | 1.535 | 1.05 |
| $\delta N_a$ | 0.15 | 0.20 | 0.15 | 0.46 | 0.67 | 0.75 |
| $\delta N_e$ | 0.85 | 0.80 | 0.85 | 0.54 | 0.33 | 0.25 |
| $\varepsilon_{host}(\lambda_a)$ | 2.624 | 2.624 | 2.624 | 2.624 | 2.624 | 2.624 |
| $\varepsilon_{real}(\lambda_a)$ | 2.624 | 2.624 | 2.624 | 2.624 | 2.624 | 2.624 |
| $\varepsilon_{imag}(\lambda_a)$ | 0.607 | 0.370 | 0.614 | 0.603 | 1.407 | 1.009 |
| $n_{host}(\lambda_a)$ | 1.62 | 1.62 | 1.62 | 1.62 | 1.62 | 1.62 |
| $n(\lambda_a)$ | 1.631 | 1.624 | 1.631 | 1.631 | 1.674 | 1.649 |
| $\kappa(\lambda_a)$ | 0.186 | 0.114 | 0.188 | 0.185 | 0.420 | 0.306 |
| $\varepsilon_{host}(\lambda_e)$ | 2.624 | 2.624 | 2.624 | 2.624 | 2.624 | 2.624 |
| $\varepsilon_{real}(\lambda_e)$ | 2.657 | 2.650 | 2.658 | 2.704 | 2.860 | 2.805 |
| $\varepsilon_{imag}(\lambda_e)$ | -0.381 | -0.238 | -0.631 | -0.263 | -0.374 | -0.188 |
| $\varepsilon_{thr}$ | -0.393 | -0.165 | -0.393 | -0.165 | -0.393 | -0.165 |
| $\varepsilon_{imag}(\lambda_e)/\varepsilon_{thr}$ | 0.97 | 1.44 | 1.61 | 1.59 | 0.95 | 1.14 |
| $n_{host}(\lambda_e)$ | 1.62 | 1.62 | 1.62 | 1.62 | 1.62 | 1.62 |
| $n(\lambda_e)$ | 1.634 | 1.630 | 1.642 | 1.646 | 1.695 | 1.646 |
| $\kappa(\lambda_e)$ | -0.117 | -0.073 | -0.192 | -0.080 | -0.110 | -0.056 |

**Table S2.** The dye concentration, population inversion and the dielectric properties of the inspected GMD and GMG nanoresonators. The total dye concentration (*N*), the population differences *($\delta N_a = \Delta N_a/N$; $\delta N_e = \Delta N_e/N$)*, the permittivity and refractive index of the host medium ($\varepsilon_{host}$, $n_{host}$), the real and imaginary parts of the permittivity and refractive index of the active systems ($\varepsilon_{real}$, $\varepsilon_{imag}$, *n*, *κ*), the analytically derived spaser threshold ($\varepsilon_{thr}$) and the ratio of the gain and threshold permittivity ($\varepsilon_{imag}(\lambda_e)/\varepsilon_{thr}$). The absorption and emission wavelength is indicated by $\lambda_a$ and $\lambda_e$.

| local **E**-field | | | | | | | | |
|---|---|---|---|---|---|---|---|---|
| | passive | | NF-type | | FF-type | | ECST-type | |
| | GMD | GMG | GMD | GMG | GMD | GMG | GMD | GMG |
| $E_{pump}$ (*$10^5$ V/m) | - | - | 2.9 | 3.5 | 2.9 | 2.2 | 1.32 | 1.52 |
| $NFE_{pump}$ | - | - | 17.2 | 13.1 | 17.1 | 20.9 | 11.1 | 8.0 |
| $E_{probe}$ (*$10^4$ V/m) | 1 | 1 | 1 | 1 | 1 | 1 | 1 | 1 |
| $E_{avg}$ (*$10^6$ V/m) | 0.48 | 0.38 | 2.76 | 2.34 | 1.16 | 1.47 | 0.85 | 0.60 |
| $E_{max}$ (*$10^6$ V/m) | 0.49 | 0.65 | 2.81 | 3.92 | 1.18 | 2.40 | 0.87 | 0.94 |
| $NFE_{avg\_probe}$ | 48.4 | 38.3 | 276 | 234 | 116 | 147 | 85.4 | 60.1 |
| $NFE_{max\_probe}$ | 49.3 | 64.6 | 281 | 392 | 118 | 240 | 87.1 | 93.7 |

**Table S3.** The near-field enhancements of the inspected GMD and GMG nanoresonators. The pump **E**-field strength *($E_{pump}$)*, the average pump **E**-field enhancement *($NFE_{pump}$)*, the probe **E**-field amplitude *($E_{probe}$)* the average and maximal local **E**-filed *($E_{avg}$ & $E_{max}$)* and the average and maximal near-field enhancement at the lasing wavelength *($NFE_{avg\_probe}$, $NFE_{max\_probe}$)*.

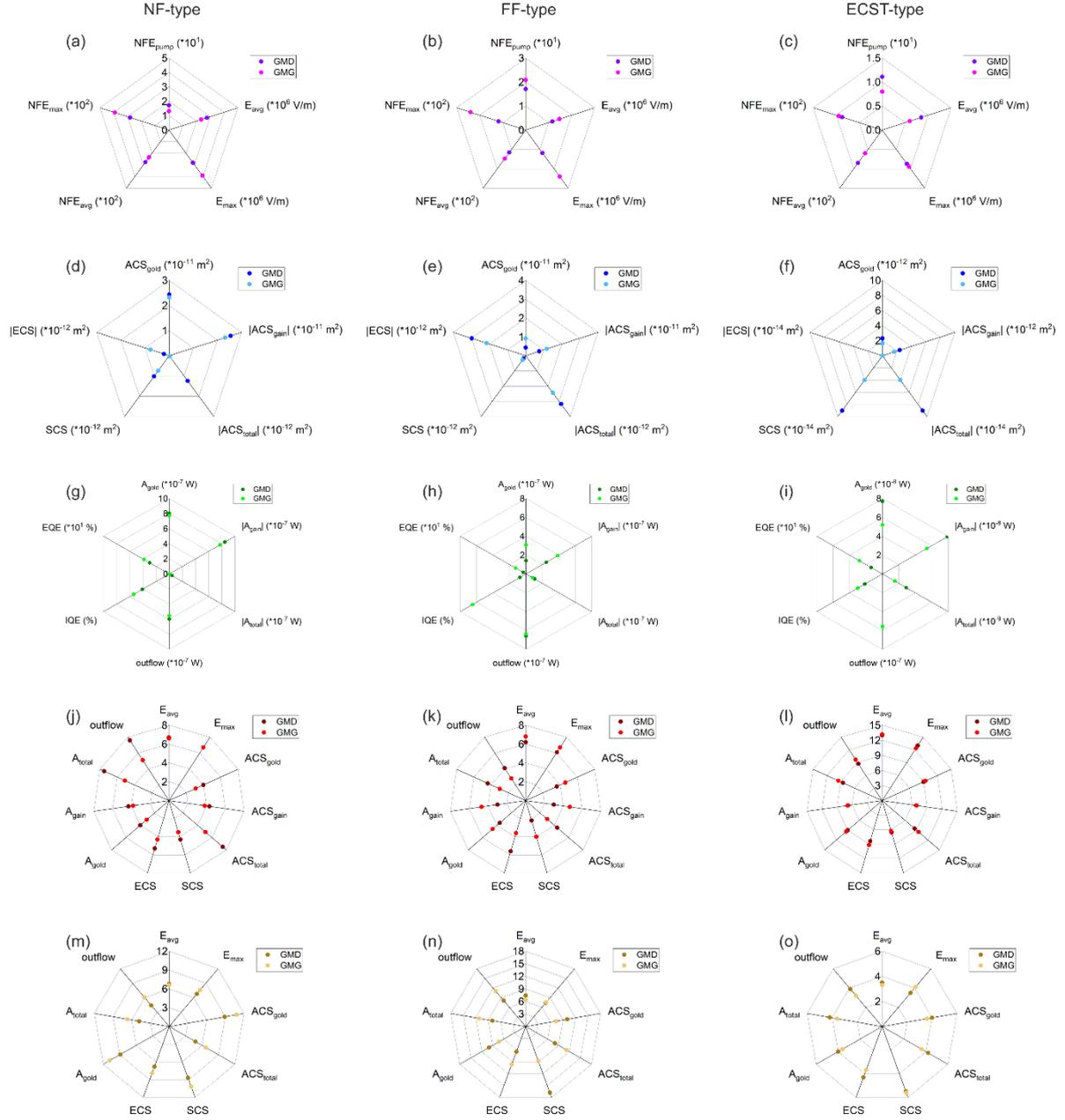

**Figure S3.** Ranking of the inspected nanoresonators. The ranking of the (a-c) near-field, the (d-f) optical cross section, the (g-i) optical response, the (j-l) FWHM and the (m-o) degree of narrowing in case of the (a, d, g, j, m) NF-type, (b, e, h, k, n) FF-type and (c, f, I, l, o) ECST-type NRs.

| | Optical cross sections | | | | | | | |
|---|---|---|---|---|---|---|---|---|
| | passive | | NF-type | | FF-type | | ECST-type | |
| | GMD | GMG | GMD | GMG | GMD | GMG | GMD | GMG |
| $ACS_{gold}$ (*$10^{-11}$ $m^2$) | 0.075 | 0.062 | 2.43 | 2.32 | 0.43 | 0.92 | 0.23 | 0.16 |
| $ACS_{gain}$ (*$10^{-11}$ $m^2$) | - | - | -2.55 | -2.32 | -0.75 | -1.17 | -0.24 | -0.16 |
| $ACS_{total}$ (*$10^{-12}$ $m^2$) | 0.075 | 0.062 | -1.24 | 0.04 | -3.17 | -2.44 | -0.09 | -0.04 |
| $SCS$ (*$10^{-12}$ $m^2$) | 0.003 | 0.002 | 1.02 | 0.74 | 0.18 | 0.28 | 0.09 | 0.04 |
| $ECS$ (*$10^{-12}$ $m^2$) | 0.078 | 0.064 | -0.22 | 0.78 | -2.99 | -2.16 | $1.1*10^{-5}$ | $2*10^{-6}$ |

**Table S4.** The optical cross section of the inspected GMD and GMG nanoresonators. The absorption cross-section of the gold ($ACS_{gold}$) and gain ($ACS_{gain}$) segments and the total absorption cross-section ($ACS_{total}$), the scattering ($SCS$) and extinction ($ECS$) cross-section, and the ratio of the $ACS$ and $SCS$ ($|ACS|/SCS$).

| Optical response, quantum efficiency and quadrupole ratio | | | | | | | | |
|---|---|---|---|---|---|---|---|---|
| | passive | | NF-type | | FF-type | | ECST-type | |
| | GMD | GMG | GMD | GMG | GMD | GMG | GMD | GMG |
| $A_{gold}$ (*10$^{-7}$ W) | 0.25 | 0.20 | 8.06 | 7.71 | 1.42 | 3.07 | 0.77 | 0.52 |
| $A_{gain}$ (*10$^{-7}$ W) | - | - | -8.47 | -7.70 | -2.48 | -3.88 | -0.79 | -0.54 |
| $A_{total}$ (*10$^{-7}$ W) | 0.25 | 0.20 | -0.41 | 0.01 | -1.06 | -0.81 | -0.029 | -0.015 |
| out (*10$^{-7}$ W) | 5.30 | 5.35 | 5.98 | 5.55 | 6.60 | 6.37 | 5.58 | 5.57 |
| $r_{out\_in}$ | 0.956 | 0.963 | 1.077 | 1.000 | 1.189 | 1.148 | 1.005 | 1.004 |
| IQE (%) | - | - | 4.07 | 5.41 | 0.74 | 6.48 | 2.14 | 3.01 |
| EQE (%) | - | - | 2.97 | 3.83 | 3.38 | 12.67 | 13.72 | 27.87 |
| QP:DP ratio (710 nm) (%) | 2.78 | 2.22 | 1.11 | 0.28 | 1.11 | 0.56 | 2.5 | 1.67 |

**Table S5.** The optical responses of the inspected GMD and GMG nanoresonators. The absorption inside the gold ($A_{gold}$) and gain ($A_{gain}$) segments and the total absorption ($A_{total}$); the power outflow (*out*) and the outflow ratio compared to the probe power ($r_{out\_in}$); the internal (*IQE*) and external (*EQE*) quantum efficiency; the quadrupole-to-dipole (*QP:DP*) ratio.

| Full width at half maximum (nm) | | | | | | | | |
|---|---|---|---|---|---|---|---|---|
| | passive | | NF-type | | FF-type | | ECST-type | |
| | GMD | GMG | GMD | GMG | GMD | GMG | GMD | GMG |
| $E_{avg}$ | 45.5 | 43.3 | 6.7 | 6.6 | 6.2 | 6.8 | 13.0 | 13.2 |
| $E_{max}$ | 45.5 | 51.1 | 6.7 | 6.7 | 6.1 | 6.7 | 13.0 | 12.4 |
| $ACS_{gold}$ | 35.7 | 34.0 | 4.0 | 3.1 | 3.6 | 4.6 | 9.0 | 9.4 |
| $ACS_{gain}$ | - | - | 4.3 | 3.8 | 3.0 | 4.7 | 6.8 | 6.9 |
| $ACS_{total}$ | 35.7 | 34.0 | 7.5 | 5.1 | 4.4 | 3.0 | 8.5 | 9.5 |
| SCS | 36.2 | 35.4 | 4.3 | 3.5 | 2.2 | 4.0 | 6.6 | 6.3 |
| ECS | 35.7 | 34.0 | 5.3 | 4.3 | 5.6 | 3.6 | 8.4 | 9.2 |
| $A_{gold}$ | 35.7 | 34.0 | 4.0 | 3.1 | 3.6 | 4.6 | 9.0 | 9.4 |
| $A_{gain}$ | - | - | 4.3 | 3.8 | 3.0 | 4.7 | 6.8 | 6.9 |
| $A_{total}$ | 35.7 | 34.0 | 7.5 | 5.1 | 4.4 | 3.0 | 8.5 | 9.5 |
| out | 33.5 | 31.3 | 7.6 | 5.1 | 4.1 | 2.8 | 8.7 | 9.7 |

**Table S6.** Full width at half maximum values in the passive and active GMD and GMG nanoresonators. The *FWHM* of the average ($E_{avg}$) and maximal ($E_{max}$) local **E**-field, the gold ($ACS_{gold}$), gain ($ACS_{gain}$) and total ($ACS_{total}$) absorption cross-section, the scattering (*SCS*) and extinction (*ECS*) cross-sections, the gold ($A_{gold}$), gain ($A_{gain}$) and total absorption ($A_{total}$) the outflow (*out*).

| Degree of narrowing | | | | | | |
|---|---|---|---|---|---|---|
| | NF-type | | FF-type | | ECST-type | |
| | GMD | GMG | GMD | GMG | GMD | GMG |
| $E_{avg}$ | 6.8 | 6.6 | 7.4 | 6.4 | 3.5 | 3.3 |
| $E_{max}$ | 6.8 | 7.6 | 7.4 | 7.6 | 3.5 | 4.1 |
| $ACS_{gold}$ | 8.9 | 10.8 | 10.0 | 7.3 | 4.0 | 3.6 |
| $ACS_{total}$ | 4.8 | 6.7 | 8.0 | 11.2 | 4.2 | 3.6 |
| SCS | 8.7 | 10.2 | 16.8 | 8.8 | 5.5 | 5.6 |
| ECS | 6.8 | 7.9 | 6.4 | 9.6 | 4.3 | 3.7 |
| $A_{gold}$ | 8.9 | 10.8 | 10.0 | 7.3 | 4.0 | 3.6 |
| $A_{total}$ | 4.8 | 6.7 | 8.0 | 11.2 | 4.2 | 3.6 |
| out | 4.4 | 6.1 | 8.1 | 11.0 | 3.9 | 3.2 |

**Table S7.** The degree of narrowing compared to the passive GMD and GMG nanoresonators. The narrowing of the average ($E_{avg}$) and maximal ($E_{max}$) local **E**-field, the gold ($ACS_{gold}$) and total absorption cross-section ($ACS_{total}$), the scattering (*SCS*) and extinction (*ECS*) cross-sections, the gold ($A_{gold}$) and total absorption ($A_{total}$), the outflow (*out*).